\documentclass[12pt,preprint]{aastex7}
\usepackage{graphicx} % Required for inserting images
\usepackage{natbib}
\usepackage{comment}
\usepackage{xspace}

\newcommand{\Fermilat}{\textit{Fermi}-LAT\xspace}

\shorttitle{FL16Y}
\shortauthors{\Fermilat collaboration}

\begin{document}

\title{\Fermilat 16-year Source List}

\author{J.~Ballet}
\email[show]{Jean.Ballet@cea.fr}
\affiliation{Universit\'e Paris Saclay and Universit\'e Paris Cit\'e, CEA, CNRS, AIM, F-91191 Gif-sur-Yvette, France}
\author{P.~Bruel}
\email[show]{Philippe.Bruel@llr.in2p3.fr}
\affiliation{LLR, CNRS/IN2P3, \'Ecole polytechnique, Institut Polytechnique de Paris, F-91128 Palaiseau, France}
\author{T.~H.~Burnett}
\email[show]{tburnett@u.washington.edu}
\affiliation{Department of Physics, University of Washington, Seattle, WA 98195-1560, USA}
\author{B.~Lott}
\email[show]{lott@cenbg.in2p3.fr}
\affiliation{Universit\'e Bordeaux, CNRS, LP2I Bordeaux, UMR 5797, F-33170 Gradignan, France}
\collaboration{5}{The \Fermilat collaboration}

\begin{abstract}
The current Fermi-LAT source catalog (4FGL-DR4: 7194 sources over 14 years) was built incrementally from the 8-year catalog. In a survey mission like Fermi, data accumulate on each source over time, so after 16 years (reached in August 2024) and twice the data for the original 4FGL sources we have more precise localization (by 24\% on average). It is thus time to generate a new original catalog, which implies, beyond adding the sources newly detectable after two more years, changing the existing source names (derived from their coordinates) and reviewing the associations.

We present an early 16-year list (FL16Y) of 7220 sources, which relocalizes all sources and improves a few aspects of the catalog analysis, but still uses the same model of interstellar diffuse emission as 4FGL-DR4.
\end{abstract}

\keywords{ Gamma rays: general --- surveys --- catalogs}

\section{Introduction}

The current Fermi-LAT source catalog\footnote{See \url{https://fermi.gsfc.nasa.gov/ssc/data/access/lat/14yr_catalog/}.} \citep[4FGL-DR4:][]{LAT22_4FGLDR3, LAT24_4FGLDR4} was built incrementally from the 8-year catalog \citep{LAT20_4FGL}. This means that, except for a few specific cases, the positions of the original (DR1) sources (and their error ellipses) were frozen to what they were after eight years. Among the 7194 DR4 sources, nearly 600 had moved outside their 95\% error radius when considering the new seed localization, beyond the 5\% expected by chance.  Furthermore, about 300 were inherited from former 4FGL catalogs but formally below the detection threshold. So we considered that the concept of an incremental catalog reached its limit, and that we had to build a fresh one.

% The major limitation of the current LAT catalogs is the precision of the model of interstellar diffuse emission. It is estimated at 3\%, which is good but not enough for the statistical precision reached below 1 GeV, particularly in the Galactic Ridge. Therefore a full 5FGL catalog covering 16 years of data requires a new model of interstellar diffuse emission, which is not yet available. In the meantime we provide an interim source list, that we call FL16Y (for Fermi-LAT 16-year), using the currently available model of interstellar diffuse emission. Its main purpose is to follow blazar activity and improve the catalog depth outside the Galactic plane.
We need to follow blazar activity and continue improving the LAT catalog depth outside the Galactic plane. For that purpose we provide an interim source list, that we call FL16Y (for Fermi-LAT 16-year)\footnote{Accessible at \url{https://fermi.gsfc.nasa.gov/ssc/data/access/lat/fl16y/}.}. Close to the Galactic plane the major limitation remains the precision of the model of interstellar diffuse emission. It is estimated at 3\%, which is good but not enough for the statistical precision reached below 1 GeV, particularly in the Galactic Ridge. As for all the 4FGL  series of catalogs, we mitigate that systematic effect with weights that attenuate the bright areas at low energies, ensuring that the quality of FL16Y in the plane is at least as good as 4FGL-DR4. Users wishing to analyze in more detail areas close to the Galactic plane must use weights, or develop a dedicated model of the local interstellar emission. The LAT collaboration is preparing a new model of interstellar emission, which will eventually allow generating a 5FGL catalog, more reliable close to the Galactic plane.

Section \ref{seds} explains specific changes to the spectral energy distributions, Section \ref{seeds} describes the construction of that list and Section \ref{assocs} explains the changes in the association process.

\section{Spectral energy distributions}
\label{seds}

The fluxes in eight individual bands for the Spectral Energy Distributions (SED) are obtained by freezing the power-law index to that obtained in the fit over the full range and adjusting the normalization in each spectral band \citep{LAT20_4FGL}. For the curved spectra, LogParabola and PLSuperExpCutoff4 (PLEC4), the photon index in a band is set to the local spectral slope at the logarithmic mid-point of the band $[E_n,E_{n+1}]$, restricted to be in the interval [0,5].

\subsection{Reason for changing}

The SED generation is the only part of the analysis that changed significantly since the DR4 catalog.
The previous approach suffered from serious limitations at low energy (in the first two bands mostly, below 300 MeV). 
The problems at low energy are due to the very broad PSF (68\% containment radius of $2.2\degr$ at 300 MeV, $8.7\degr$ at 50 MeV). It leads to a low source-to-background ratio and large confusion (the average distance between sources is less than $1.5\degr$). It made iterations in those bands very slow and unstable, and sometimes resulted in failures due to convergence errors.
The previous method assumed nothing about the source fluxes (i.e., each band was treated independently). This ensured absolute independence between the SED points. But it meant that nothing prevented a faint source with a relatively hard spectrum from becoming formally bright at low energy (with very large error bars). This is of course not prevented by any physical law (very soft sources with a high energy tail can exist) but attributing that flux to an adjacent brighter softer source would be a simpler physical solution.

\subsection{Methodology}

The idea of the new approach is to use information from the global spectral fit (over all energies) to set (Gaussian) priors when fitting the SED points, penalizing large excursions away from the global model. The prior mean in each individual band can be naturally defined as the model prediction integrated over that band.
Defining a prior sigma, however, requires careful attention in order not to overconstrain the SED fits, rendering the SED points useless as a check of the global spectral fit. To that end, the prior sigma $\sigma_P$ is obtained as a quadratic combination of four terms, so that the largest one dominates.
\begin{equation}
\sigma_P = \sqrt{\alpha^2 \sigma_C^2 + \alpha^2 \sigma_S^2 + \sigma_M^2 F_M^2 + F_M^2}
\label{eq:priordef}
\end{equation}
The first term $\sigma_C$ is the covariance error, directly obtained from propagating the covariance matrix of the parameters of the global spectral model. The second term $\sigma_S$ is the statistical error, what one would get from Poisson statistics (including likelihood weights) if the source was alone on a flat background. It is normally smaller than $\sigma_C$ at low energy, but acts as a lower limit when the model prediction is very low (very curved model fit). The first two terms are multiplied by a coefficient $\alpha$, currently set to 3, so that the prior is first felt at about 3$\sigma$. The last two terms are proportional to the flux $F_M$ predicted by the global model. The third term $\sigma_M$ is the model error. It applies only to the power-law fits, which have only one spectral parameter and therefore smaller covariance errors than the LogParabola fits. It is obtained in relative terms as $\sigma_M = \ln(F_{LP}/F_M)$, where $F_{LP}$ is the model prediction of the best LogParabola fit. Finally, the fourth term is just the model prediction, so that the relative error on the final SED point, if the data are not constraining at all, is at least one.
The result appears in the \texttt{PriorSigma\_Band} vector column of the FITS file.
We do not apply the priors above 10 GeV, in a range where the PSF is narrow enough that the background is low and the confusion small.

\subsection{Application to the DR4 catalog}

\begin{figure}[!ht]
   \centering
   \begin{tabular}{cc}
   \includegraphics[width=0.49\textwidth]{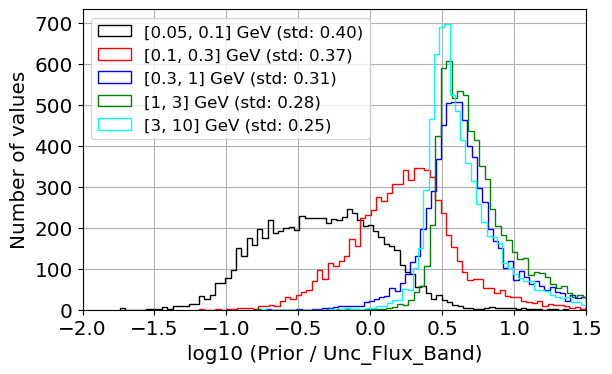} & 
   \includegraphics[width=0.49\textwidth]{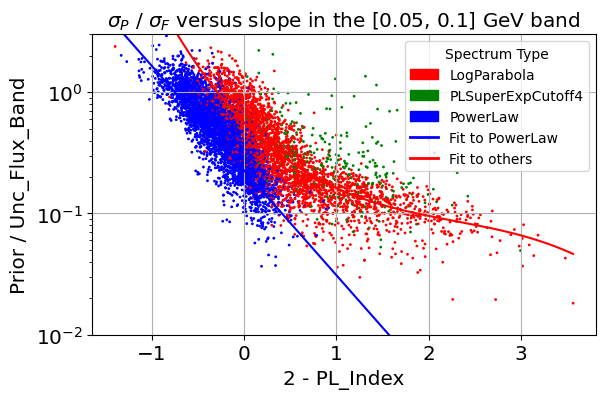}
   \end{tabular}
   \caption{Prior sigma $\sigma_P$ for band fluxes compared to the errors in the SED calculation without priors, on DR4 sources. Left: Distributions of the ratio between $\sigma_P$ and SED error in all bands. The average ratio increases from band 1 to band 4 (best constraints), and decreases again in higher-energy bands. It is less than one (so the prior has a strong impact) in many sources in bands 1 and 2 only. The numbers in parentheses are the widths (standard deviation) of the distributions (in log$_{10}$, like the abscissa). Right: Correlation between that ratio and the spectral index in band 1 only, separately for each spectral type used in the model. For LogParabola and PLEC4, 2$-$PL\_Index is replaced by the average index in $\nu F_\nu$ between band 1 and the pivot frequency $\nu_0$, i.e. $1+\ln(F_\nu(\nu_1)/F_\nu(\nu_0))/\ln(\nu_1/\nu_0)$, where $\nu_1$ is the logarithmic mid-point of the band. That quantity (intermediate between 2$-$PL\_Index itself - appropriate at $\nu_0$ - and the local spectral slope at $\nu_1$) resulted in the narrowest correlation. The ratio decreases from soft to hard sources in the first band, because hard sources are constrained by higher energies.}
\label{fig:priortosigma}
\end{figure}

\begin{figure}[!ht]
   \centering
   \begin{tabular}{cc}
   \includegraphics[width=0.49\textwidth]{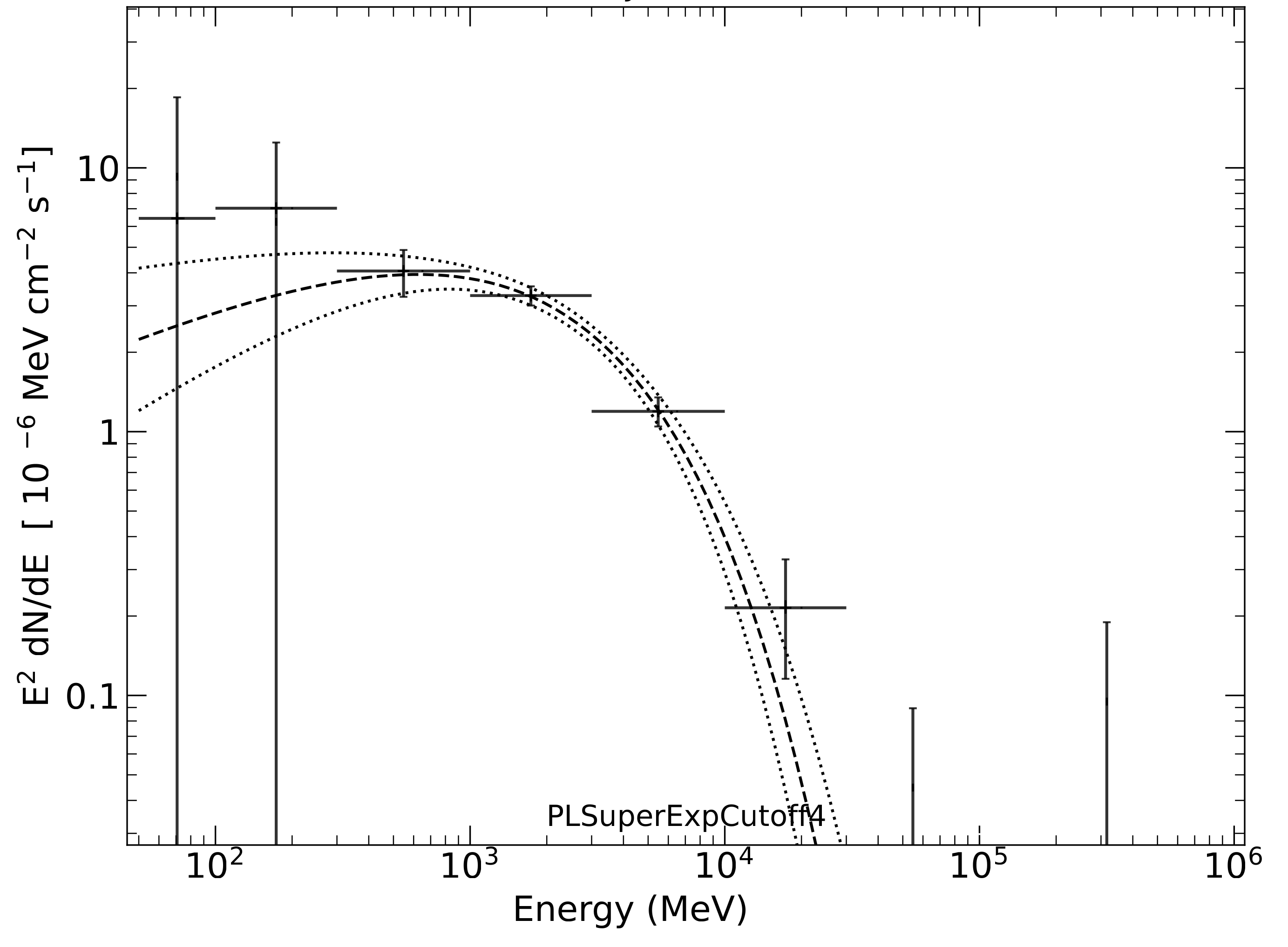} & 
   \includegraphics[width=0.49\textwidth]{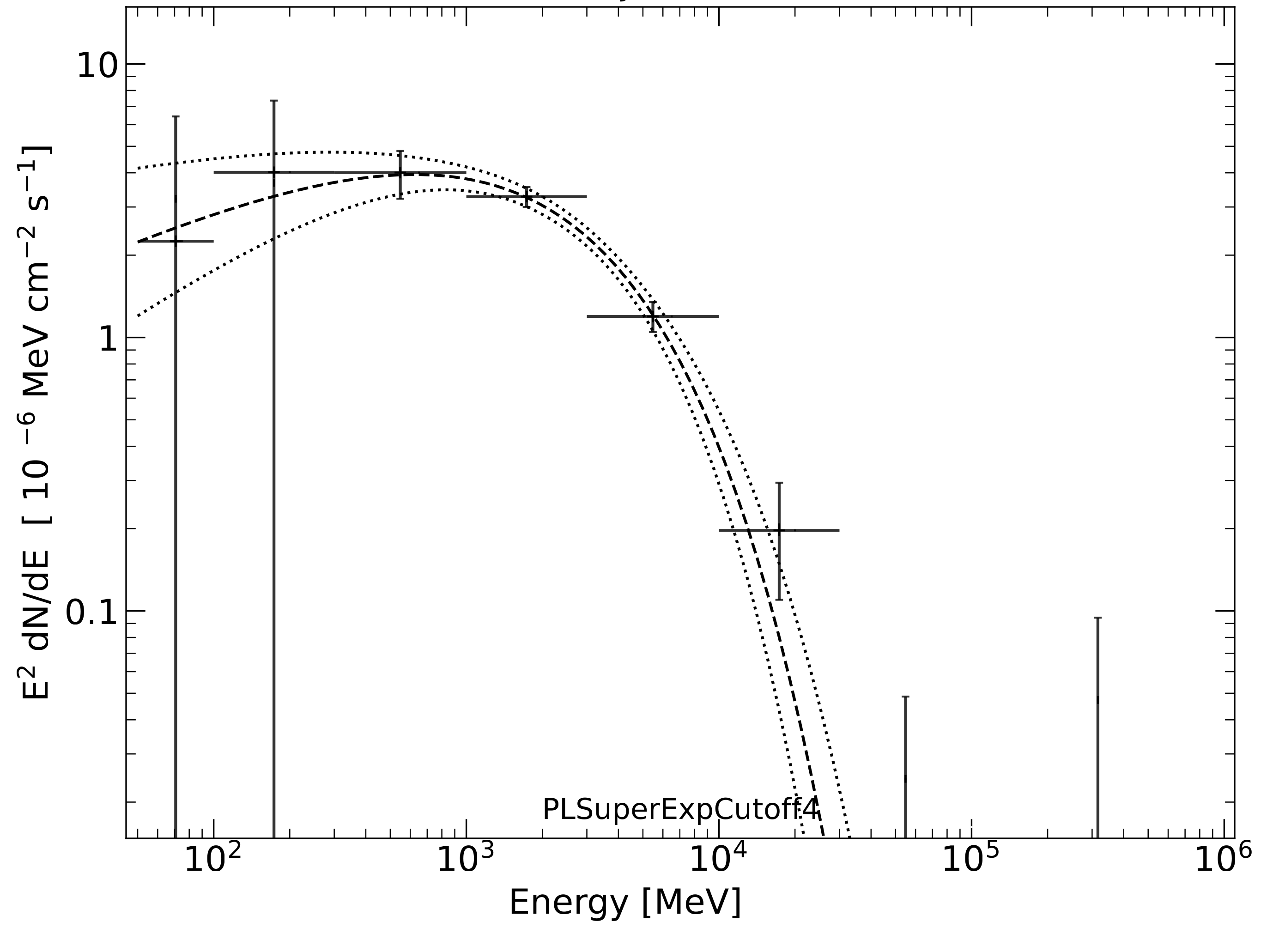}
   \end{tabular}
   \caption{
%     SED of 4FGL J1641.0$-$4619 (HESS J1641$-$463) in the official DR4 catalog (left) and after applying priors (right).
     SED of 4FGL J1105.4$-$6108 (PSR J1105$-$6107) in the official DR4 catalog (left) and after applying priors (right).
     The effect of the priors is obvious in the first two bands. For that confused source in the Galactic plane, the very uncertain SED points in DR4 have been pulled much closer to the model prediction (dashed curve) and the errors have been reduced. The effect in the other bands (above 300 MeV) is small. The automatic Y range reacted to the differing upper limits so it is not the same in both plots. }
\label{fig:priorP88Y4309}
\end{figure}

\begin{figure}[!ht]
   \centering
   \begin{tabular}{cc}
   \includegraphics[width=0.53\textwidth]{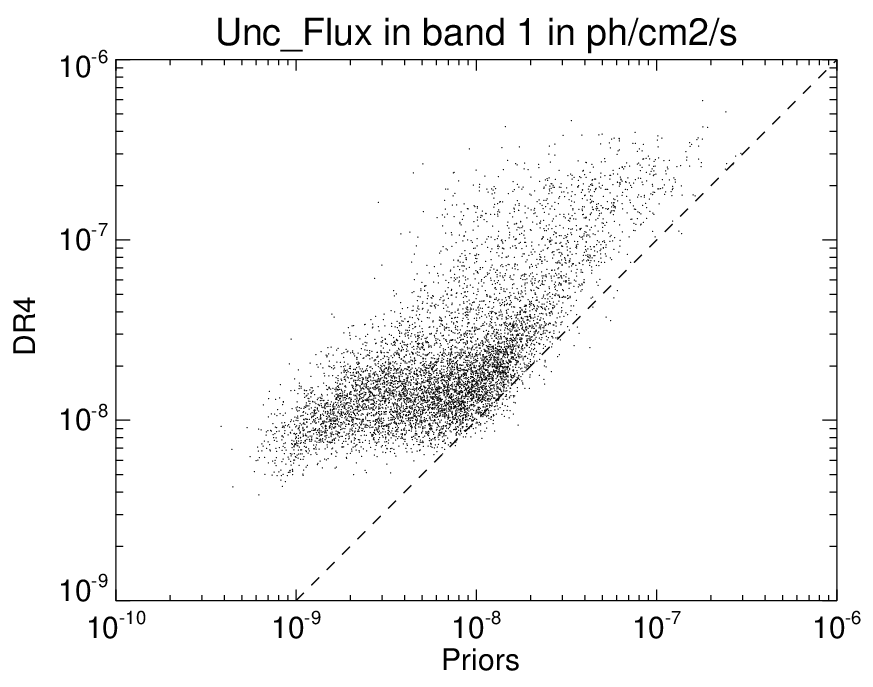} & 
   \includegraphics[width=0.45\textwidth]{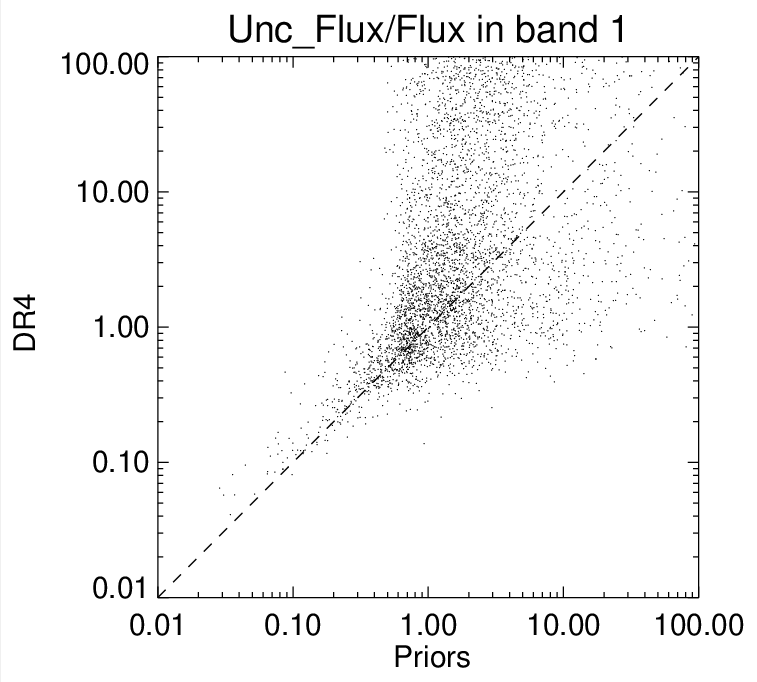}
   \end{tabular}
   \caption{Effect of the priors on the errors in band 1 (50 to 100 MeV). It is illustrated on the upper errors (second entry in \texttt{Unc\_Flux\_Band}, toward larger fluxes), because the lower errors are truncated when the best fit is close to 0. Left: Direct correlation between the errors with priors (abscissa) and the original errors (ordinate). The dashed line is the one-to-one correlation. Essentially all points are above the line, indicating that the errors are reduced. The maximal reduction is approximately a factor ten. Right: Correlation of the relative errors (with respect to the best fit). The plot looks very different because the best fit changes as well. The effect, as expected, is largest for faint sources ($dF/F > 1$), and becomes smaller for strong ones. The very strong sources ($dF/F < 0.1$) tend to be all above the one-to-one correlation, corresponding to smaller errors with priors. This behavior is not a direct consequence of the priors, since the adopted prior sigma is always at least as large as the flux. Rather, it arises indirectly from constraining nearby faint sources from deviating excessively from their global models, as intended.}
\label{fig:priorband1}
\end{figure}

Figure \ref{fig:priortosigma} (left) compares the prior sigma $\sigma_P$ obtained from Eq.\ref{eq:priordef} with the error on the band flux \texttt{Unc\_Flux\_Band} obtained without priors in DR4 (that check was done in preparation shortly after DR4). It shows that, as required, the prior will have a strong effect in band 1, less strong in band 2, and not much effect in higher-energy bands, where confusion is much less. It also shows (right) that the largest effect in band 1 is obtained for the hardest sources. The main downward trend is due to the $\sigma_C$ term. The $\sigma_S$ and $\sigma_M$ terms are what prevent the ratio from becoming even smaller, for very curved sources that appear very hard in band 1 (at right in the plot, for curved models).
Figure \ref{fig:priorP88Y4309} illustrates the effect on one particular source.
Figure \ref{fig:priorband1} illustrates the error reduction on all sources in the first band.

\section{List of seeds, thresholding and light curves}
\label{seeds}

The data for FL16Y were taken between 4 August 2008 (15:43 UTC) and 2 August 2024 (11:13 UTC), covering 16 years. During the last two years, we excised 97 ks due to six bright solar flares in October 2022, December 2023, February and July 2024 (the Sun was reaching its maximum) and 9 ks around 5 $\gamma$-ray bursts (including the very bright GRB221009A). The energy range remains 50 MeV to 1 TeV.
As before, we use the Test Statistic TS = 2 $\log (\mathcal{L} / \mathcal{L}_0)$ to quantify how significantly a source emerges from the background, comparing the maximum value of the likelihood function $\mathcal{L}$ including the source in the model with $\mathcal{L}_0$, the value without the source.

\subsection{Seeds and localization}

The list of seeds (called uw1617) was built using {\it pointlike}, as in previous catalogs. We did not update the list of extended sources, except that we used the two-template model of the Cygnus Loop \citep{CygnusLoop_Tutone21} to reduce residuals there. 
As before, we rejected seeds with bad localization, faint or unassociated seeds inside extended sources, and faint seeds overlapping another source. We also rejected a number of seeds appearing next to the Crab and Geminga pulsars, due to residuals along right ascension that are clearly not real.

Following Section 3.2 of \citet{LAT24_4FGLDR4}, we updated the list of transients to add four new faint transients from the last two years (three blazar flares and nova V1723 Sco 2024).  Those transients, like the previous ones not detected over the full time interval, have fixed power-law index (taken from their detection over one month or less). The positions of transients (old and new) present in uw1617 are left unchanged and no error ellipses are reported, unless they are better localized in uw1617 (only FL16Y J0105.9+1911, associated with the BL Lac TXS 0103+189).

We added 21 faint LAT pulsars not detected in uw1617 at their radio positions. Eight were preserved by the main pipeline and could be localized by {\it fermipy} (Section \ref{fermipy}), so we moved them in a subsequent iteration to their best $\gamma$-ray positions and they appear in the final catalog as ordinary sources with an error ellipse.

The all-sky verification led to adding nine point sources inside extended sources to account for strong residuals at TS $>$ 100 (in particular one in W 51C).
They correspond to uw1617 seeds so they have a proper localization (except FL16Y J2033.0+3900, inherited from 4FGL).
%, and three fainter isolated peaks (FL16Y J0619.9+1455, J1213.8$-$1659 and J2001.4$-$0051). The former corresponded to uw1617 seeds (except FL16Y J2033.0+3900, inherited from 4FGL), and the latter were localized by {\it fermipy} (see Section \ref{fermipy}).

\begin{figure}[!ht]
   \centering
   \begin{tabular}{cc}
   \includegraphics[width=0.49\textwidth]{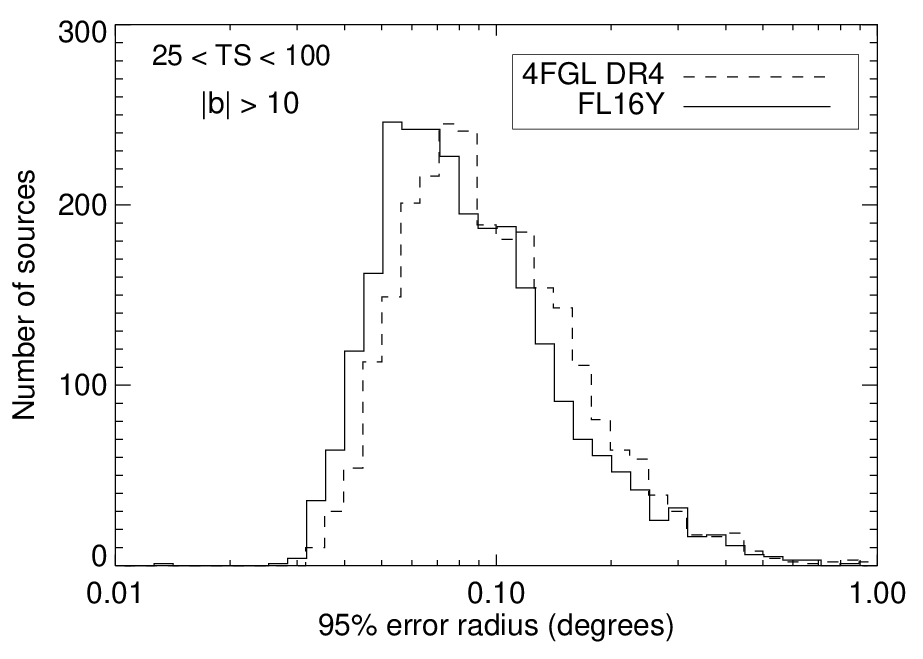} & 
   \includegraphics[width=0.49\textwidth]{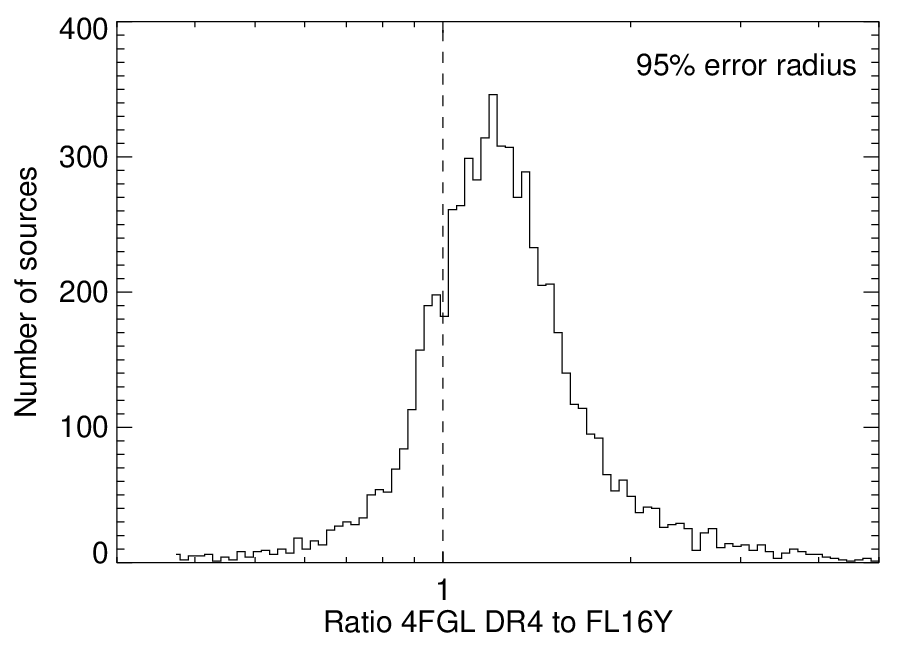}
   \end{tabular}
   \caption{Localization improvement between 4FGL and FL16Y. Left: Error radius distribution in FL16Y (solid) compared to 4FGL-DR4 (dashed), restricted to sources at low TS and high latitude (comparable). The means are $4.6\arcmin$ and $5.3\arcmin$, respectively. Right: Distribution of the ratio of error radii for the same sources in DR4 and FL16Y. The mean is 1.24. }
\label{fig:localization}
\end{figure}

The total number of point-like seeds was about 11,800, among which nearly 6,900 had a 4FGL counterpart (using the simple matching criterion $\theta < \sqrt{\theta_{95}^2(DR4) + \theta_{95}^2(FL16Y)}$, where $\theta$ is the angular distance between the sources in the two catalogs, and $\theta_{95}$ is \texttt{Conf\_95\_SemiMajor}).
Figure \ref{fig:localization} (left) shows that localization of faint sources at high latitudes (a relatively uniform sample) improves between 4FGL and FL16Y. {This plot compares sources at the same TS $<$ 100, but going deeper provides more high-energy photons that carry the best angular information. Figure \ref{fig:localization} (right) shows how the localization of individual sources improved (because the exposure was multiplied by a factor two for all sources detected in DR1). A factor of 1.24 means a reduction by 1/3 of the area of the error ellipse. The large scatter in the distribution is due to source variability (some sources just got fainter in recent years) and additional confusion (there are 52\% more sources in FL16Y than DR1).

\subsection{Spectral shapes}
\label{shapes}

We freed the exponential index in six more pulsars, and switched two bright blazars (Ton 599 and BL Lac) from LogParabola to PLEC4 with free exponential index.

\begin{figure}[!ht]
   \centering
   \begin{tabular}{ccc}
   \includegraphics[width=0.32\textwidth]{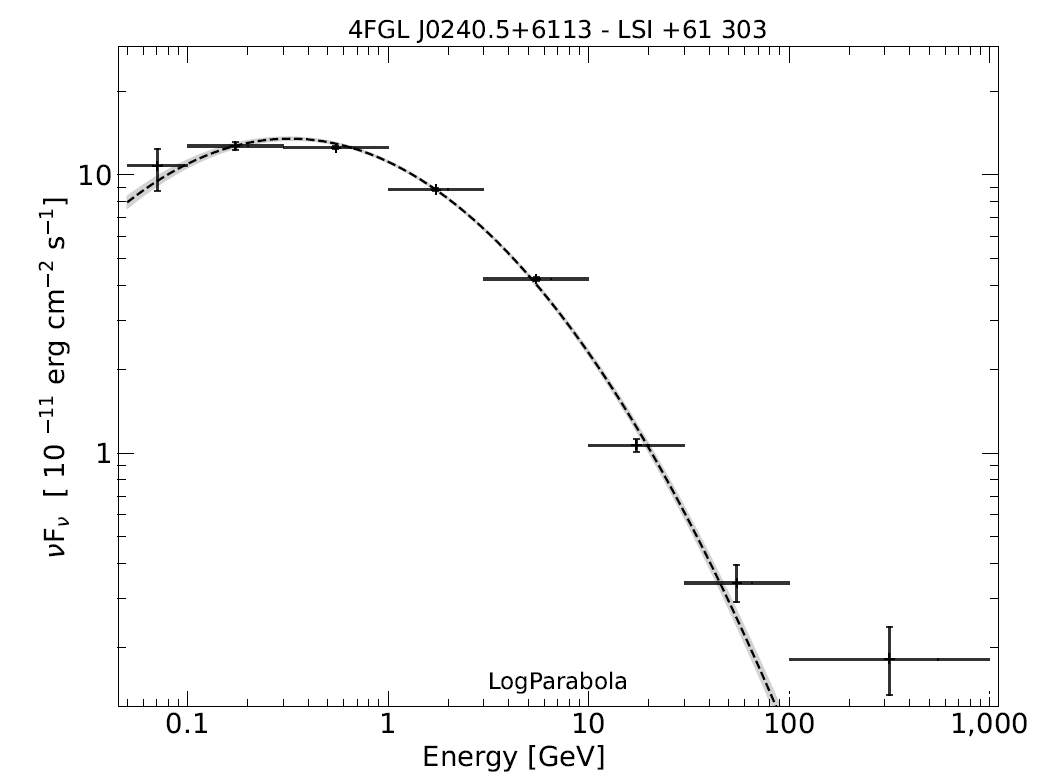} & 
   \includegraphics[width=0.32\textwidth]{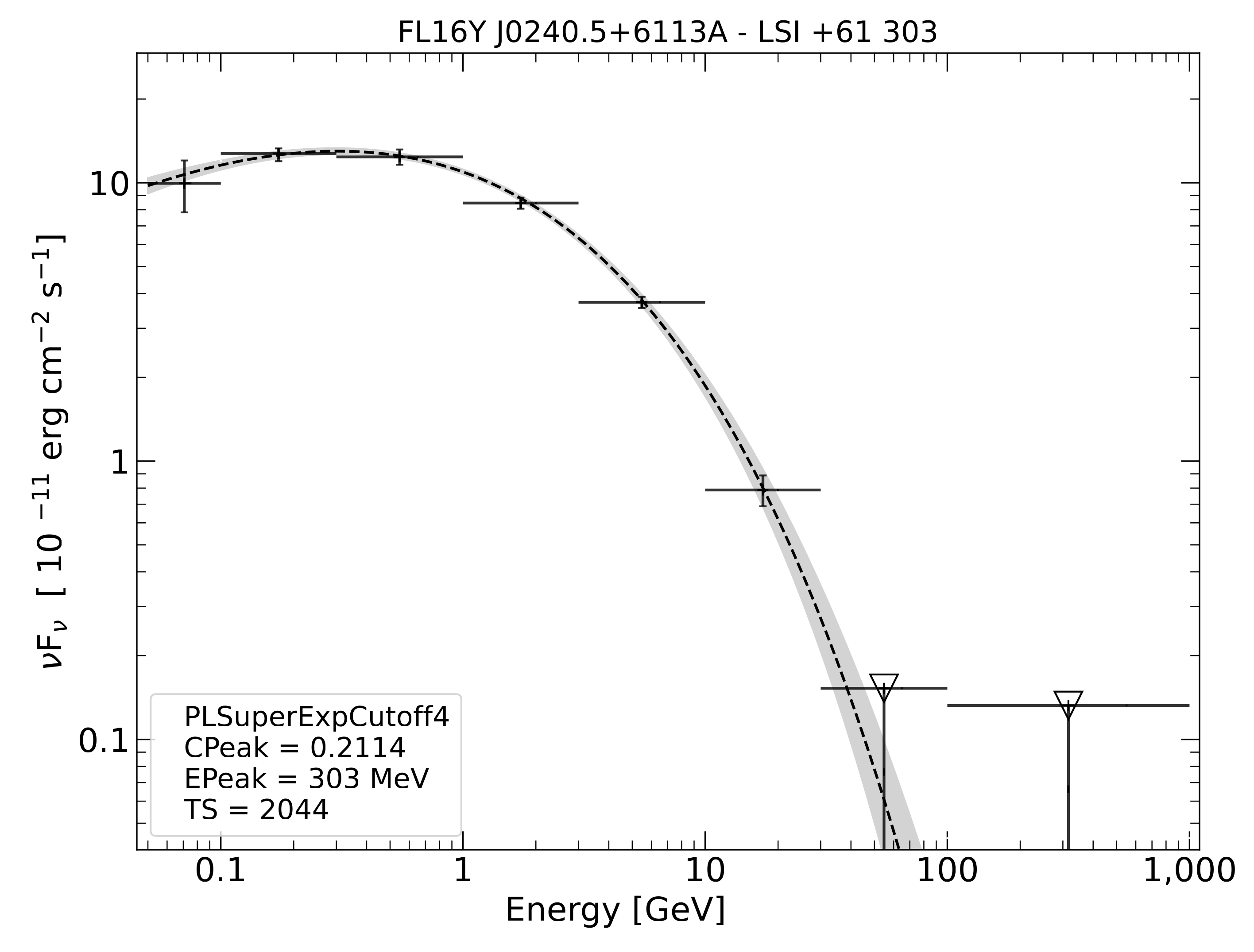}
   \includegraphics[width=0.32\textwidth]{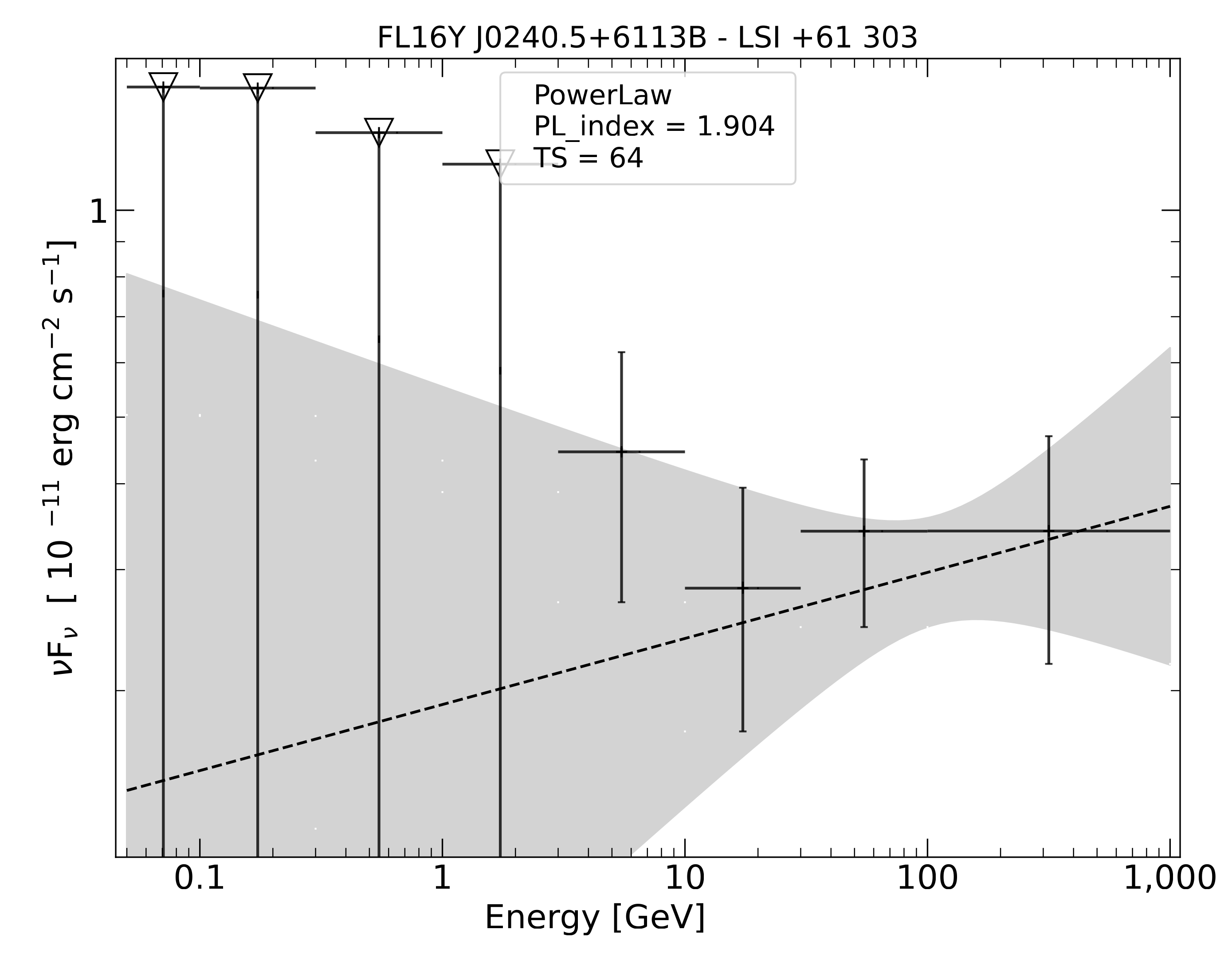}
   \end{tabular}
   \caption{Spectral fit improvement between 4FGL DR4 and FL16Y for LS I +61 303 at TS $\simeq$ 71,000 as a single source. Left: Spectral fit with a single LogParabola in DR4. Center: Spectral fit of the main (soft) component with PLEC4 in FL16Y. Right: Spectral fit of the secondary (hard) component with a power law in FL16Y. The global TS in FL16Y increased by 42 between the two models.}
\label{fig:lsi+61}
\end{figure}

\begin{figure}[!ht]
   \centering
   \begin{tabular}{cc}
   \includegraphics[width=0.49\textwidth]{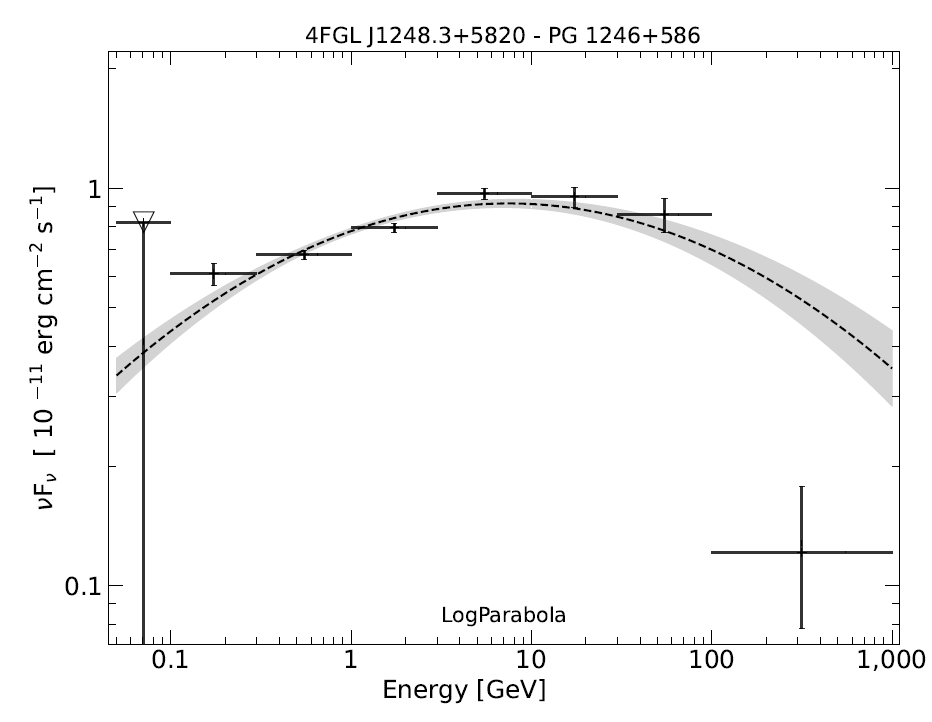} & 
   \includegraphics[width=0.49\textwidth]{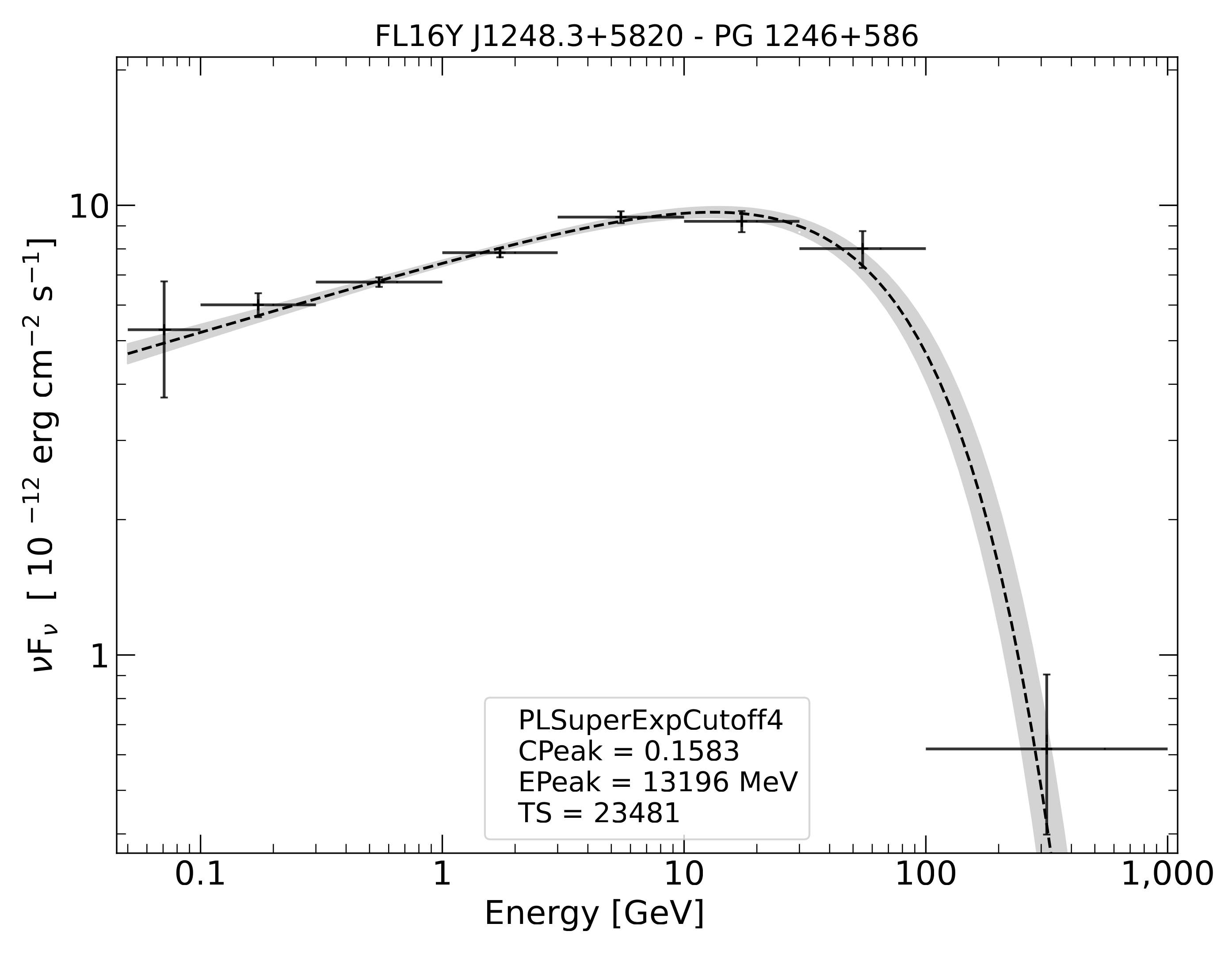}
   \end{tabular}
   \caption{Spectral fit improvement between 4FGL DR4 and FL16Y for PG 1246+586 at TS $\simeq$ 18,000 in DR4. Left: Spectral fit with a LogParabola in DR4. Right: Spectral fit with PLEC4 in FL16Y. TS in FL16Y increased by 51 between the two models.}
\label{fig:pg1246}
\end{figure}

We looked individually at all DR4 sources reported with bad spectral fit quality. This is triggered when {\texttt Spectral\_Fit\_Quality} exceeds 18.47, corresponding to $10^{-3}$ probability for 4 degrees of freedom, or when any spectral point is more than 3 $\sigma$ away from the global fit. We visually identified a number of bright sources (TS $>$ 1000) whose spectral representation could be improved, avoiding low-energy residuals that could affect neighbors. They came in two categories:
\begin{enumerate}
\item Three sources could not be represented well by a single component: the $\gamma$-ray binaries LS I +61 303 and LS 5039, and the core of the Cen A radio galaxy. We added a power-law spectral component to all three (see Figure \ref{fig:lsi+61}), after fitting the (curved) main component. For those three the additional power law was both formally significant and stable (little correlation with the main component). We also switched the main component of the two binaries to PLEC4 with free exponential index ($\log (\mathcal{L}_{\rm PLEC4} / \mathcal{L}_{\rm LP}) > 10$). Their spectra drop fast toward high energies, but look like power laws at low energies. The most obvious improvement was at high energy, but the most important one was at low energy, where the bad fit perturbed the neighbors. Sources split into two components appear in the catalog twice, with suffixes A and B, in order of decreasing significance.  We do not consider them to be separate sources, however, so we count 7220 sources but the catalog contains 7224 rows (the Crab nebula was split into two components, as in previous versions of the FGL catalog). A consequence of that decomposition is that those three sources have much lower TS (LS I +61 303 went down from TS $\simeq$ 71,000 in DR4 to TS = 2044 in FL16Y) than expected for their flux, just like the Crab.
\item Six other source spectra dropped fast toward high energies, but looked like power laws at low energies: five distant blazars affected by EBL absorption (B3 0133+388, TXS 0628$-$240, PG 1246+586, NVSS J141826$-$023336 and PG 1553+113), and the black-widow pulsar candidate 1FGL J0334.2+7501. We switched them from LogParabola to PLEC4 with exponential index fixed to 1 (see Figure \ref{fig:pg1246}). Again the most obvious improvement was at high energy, but the most important one was at lower energies, where the bad fits perturbed the neighbors.
\end{enumerate}

We noted that the extended source FL16Y J0851.9$-$4620e (the supernova remnant Vela Junior) was fit with an inverted LogParabola spectral shape ($\beta < 0$). This is of course unphysical and cannot be extrapolated to TeV energies, but it fits an apparently real excess below a few GeV. We attribute it to emission from the large (2.5$\degr$ radius) and old Vela supernova remnant, which is missing in our sky model, but probably has a soft spectrum similar to the Cygnus Loop. It is underlying the entire Vela region, and is probably picked up by all sources inside it. We left this inverted LogParabola in the model to avoid perturbing the entire region.

\subsection{Thresholding and light curves}

This part was done using Fermitools 2.4.0\footnote{See \url{https://fermi.gsfc.nasa.gov/ssc/data/analysis/}.}, and the same rescaled interstellar emission model as DR4.
We applied the usual cut on the Test Statistic TS = 2 $\ln (\mathcal{L} / \mathcal{L}_0) >$ 25, comparing the maximum value of the likelihood function $\mathcal{L}$ including the source in the model with $\mathcal{L}_0$, the value without the source. The threshold is raised to TS $>$ 29 for curved spectra that have one more free parameter.
Among the nearly 11,800 input point-source seeds, we kept 7220 after the {\it gtlike} pipeline. This is about 5\% more TS $>$ 25 point sources than in DR4. More than 6450 DR4 point sources (90\%) are in FL16Y, including more than 110 that had TS $<$ 25 in DR4. Only the transients were allowed to remain at TS $<$ 25 in FL16Y. Conversely, 6\% of DR4 point sources with TS $>$ 25 are not positionally compatible with FL16Y sources, among which half are close to the Galactic plane.

The number of sources with TS $>$ 25 increased since DR4 at high latitudes, as it should, but that near the Galactic plane remained approximately the same. This is not a worry, since that area of the sky is limited by the model of interstellar diffuse emission rather than statistics.
One source with TS $>$ 100 in DR4 (4FGL J1753.3$-$3325 near the Galactic Center) was split in two (although only one of the two seeds passed the final cut). Six others were shifted slightly out of the error ellipse.
Considering the DR4 sources with an obvious FL16Y counterpart, their TS increased by 12\% on average. This is a little less than the exposure increase (14\%), because more confusion reduces individual TS values, and the likelihood weights were a little reduced.

Light curves were generated as in DR4, with yearly time bins. 1981 sources are formally variable with a variability index larger than 30.58 (less than 1\% chance of being constant for 16 time bins). This corresponds to 27\% of the point sources, vs. 26\% in DR4. Most are blazars, as before.
The extended source W 51 C is formally variable, but this is due to confusion with the point source that was added inside it to remove a large residual.
Among the pulsars, those distinctly variable before remain so. PSR J1301+0833 and J1653$-$0158 are new variable candidates. The young pulsar PSR J0205+6449 is formally newly variable too, but it is superposed with its nebula 3C 58, so confusion is possible. FL16Y J0447.9$-$2751 (not a LAT pulsar, but associated with the young pulsar PSR J0448$-$2749) is formally variable, but it was associated with a blazar in DR4 so the variability may be due to confusion.  Finally, three are just above the variability threshold (compatible with fluctuations, among 300 pulsars).

\subsection{Localization check and flags}
\label{fermipy}

Up to 3FGL we checked the $pointlike$ localizations by calling $gtfindsrc$ above 1 GeV. Because $gtfindsrc$ works in unbinned mode, this became unreasonably slow after 3FGL, and no localization check was performed in the 4FGL catalogs. For 5FGL we developed an interface to the $localize$ function of $fermipy$ \citep[][v1.4.0]{fermipy17} to recover this functionality, and we applied it to FL16Y. We applied this check before the final cut at TS $>$ 25, so that faint sources (16 $<$ TS $<$ 25) were contributing to the model. The $fermipy$ localization was applied above 1 GeV only, as for $gtfindsrc$ previously, because the broad PSF and large confusion below 1 GeV imply large correlations between sources at low energies. As a result the softest sources (which do not reach TS $>$ 25 above 1 GeV) could not be checked. This affected only 168 sources.
Thanks to the $fermipy$ localization check we could revive Flag 7 (unstable position determination), which was deprecated since 3FGL \citep[Table 3 of][]{LAT15_3FGL}. 95 sources were flagged for that reason, including 33 that were not already flagged for other reasons. 

Flag 6, which was associated with the c sources, was not activated in FL16Y (it was too much work to check all sources manually).
Most c sources were already flagged for other reasons (85\% of the 354 c sources in DR4), because many flags are related to the effects of diffuse emission.

The other flags were as in DR4 \citep[Table 4 of][]{LAT24_4FGLDR4}. In particular, the flags related to the model of interstellar diffuse emission were obtained by comparing with the same analysis using the 3FGL model of interstellar diffuse emission \citep{LAT16_DiffuseModel}.
Overall 1895 entries (26\%) were flagged in FL16Y, mostly in the Galactic plane (only 13\% of entries are flagged at $|b| > 10\degr$). 

\subsection{Changes to the catalog format}

The only important change to the DR4 catalog format is the addition of the new vector column \texttt{PriorSigma\_Band} reporting the widths of the Gaussian priors used when fitting the band fluxes (see Section \ref{seds} and Eq. \ref{eq:priordef}). The mean of the Gaussian prior in each band is the flux obtained from the best spectral representation over the full band.

The \texttt{Source\_Name} column was pushed to 19 characters to accommodate the longer FL16Y prefix. The source name can now end in A or B when a source spectrum is decomposed into several components (see Section \ref{shapes}). The Crab components keep their historical names (no termination letter for the pulsar, s for the synchrotron nebula and i for the inverse Compton nebula).
No FL16Y source ends in c (see Section \ref{fermipy}).

The \texttt{DataRelease} and \texttt{ASSOC\_4FGL} columns (specific to incremental releases) were removed and the \texttt{ASSOC\_FGL} column now reports preferentially 4FGL counterparts.

\section{Associations}
\label{assocs}

We have used the same association procedure as in 4FGL-DR4, based on the Bayesian and likelihood-ratio methods. New catalogs have been considered in the Bayesian method: the 3HSP catalog of extreme and high-synchrotron peaked blazars \citep{Chan19}, the Firmamento\footnote{\url{https://firmamento.nyuad.nyu.edu/home}} compilation of blazars, the lists of supernova remnants in the LMC \citep{Zan24} and the SMC \citep{Mag19} (both treated as point-like objects), and the SpiderCat\footnote{\url{https://astro.phys.ntnu.no/SpiderCAT/}}  catalog of spider binary pulsars \citep{Kol25}. Other catalogs have been updated, including the Galactic millisecond pulsar catalog created at  West Virginia University\footnote{\url{http://astro.umd.edu/~eferrara/pulsars/GalacticMSPs.txt}}. A new survey used in the likelihood-ratio method is the eRass1 catalog \citep{Mer24} of X-ray sources in the western Galactic hemisphere.   

Table \ref{tab:classes} lists the census of FL16Y associations for the different classes. Some  2115 sources remain unassociated, representing 29\% of the total. Among the new FL16Y sources, not present in 4FGL-DR4, 284 are associated and 404 are not. The former include 227 blazars (27 FSRQs, 43 BLLs, and 157 BCUs),   22 UNKs, 24 SPPs, one PSR with detected pulsations, five MSPs (two of which with detected pulsations),  two SNRs, one globular cluster and two star-forming regions. Out of the 404 unassociated sources, 180 are located within 10$\degr$ of the Galactic plane, and 158 of them have a power-law index greater than 2.4, qualifying them as soft Galactic unassociated sources \citep[SGUs,][]{LAT22_4FGLDR3}. Concerning the sources  already present in 4FGL-DR4, 109 have lost their associations, mostly with blazars. On the other hand, 364 formerly unassociated sources  have gained an association, in part because of an improved localization (driving for example 141 new associations with blazars) and also due to the implementation of the aforementioned additional counterpart catalogs.    In addition,  77 sources have changed associations. The 3HSP catalog has enabled the classification of 249 former BCUs as BLLs. Gamma-ray pulsations have been found for 5 young pulsars and 29 millisecond pulsars. We provide low-confidence associations for 196 sources and Planck associations for 81 more.  
A peculiar association with the radio-optical astrometric reference frame star V1794 Cygni / HD 199178 \citep{Lun24}, showing strong X-ray activity,  has been obtained via the correlation with the radio fundamental catalog \citep{Petrov}. Although the association is likely to be fortuitous, we report it as the low-confidence  counterpart to  FL16Y J2053.4+4429.

\newcommand{\nAGN}{1}
\newcommand{\nPWN}{12}
\newcommand{\nspp}{128}
\newcommand{\nnlsy}{2}
\newcommand{\nsey}{4}
\newcommand{\nSEY}{0}
\newcommand{\nMSP}{167}
\newcommand{\npwn}{8}
\newcommand{\nBCU}{1}
\newcommand{\nBIN}{1}
\newcommand{\nFSRQ}{44}
\newcommand{\nSPP}{6}
\newcommand{\nmsp}{31}
\newcommand{\nbll}{1803}
\newcommand{\nGAL}{2}
\newcommand{\nHMB}{7}
\newcommand{\nbin}{22}
\newcommand{\nna}{2114}
\newcommand{\nIDna}{0}
\newcommand{\nsfr}{25}
\newcommand{\nBLL}{22}
\newcommand{\nRDG}{6}
\newcommand{\ngal}{2}
\newcommand{\nsnr}{17}
\newcommand{\nsbg}{8}
\newcommand{\nSBG}{0}
\newcommand{\npsr}{5}
\newcommand{\nlmb}{7}
\newcommand{\nUNK}{3}
\newcommand{\nLMB}{2}
\newcommand{\nhmb}{3}
\newcommand{\nGC}{1}
\newcommand{\ngc}{0}
\newcommand{\nagn}{5}
\newcommand{\nfsrq}{791}
\newcommand{\nssrq}{2}
\newcommand{\nSSRQ}{0}
\newcommand{\nglc}{36}
\newcommand{\nGLC}{0}
\newcommand{\nSFR}{3}
\newcommand{\nrdg}{75}
\newcommand{\nSNR}{26}
\newcommand{\ncss}{4}
\newcommand{\nCSS}{0}
\newcommand{\nPSR}{140}
\newcommand{\nbcu}{1512}
\newcommand{\nNOV}{7}
\newcommand{\nnov}{0}
\newcommand{\nunk}{161}
\newcommand{\nNLSY}{4}
\newcommand{\nID}{455}
\newcommand{\nass}{4651}

% Table listing the source classes and their numbers
\begin{deluxetable}{lcrcr}
\setlength{\tabcolsep}{0.04in}
\tablewidth{0pt}
\tabletypesize{\scriptsize}
\tablecaption{LAT FL16Y Source Classes 
\label{tab:classes}
}
\tablehead{
\colhead{Description} & 
\multicolumn{2}{c}{Identified} &
\multicolumn{2}{c}{Associated} \\
& 
\colhead{Designator} &
\colhead{Number} &
\colhead{Designator} &
\colhead{Number}
}
\startdata
Young pulsars, identified by pulsations  &  PSR  & \nPSR &  \nodata  & \nodata \\ 
Young pulsars, no pulsations seen in LAT yet  &  \nodata  & \nodata &  psr  & \npsr \\ 
Millisecond pulsars, identified by pulsations  &  MSP  & \nMSP &  \nodata  & \nodata \\ 
Millisecond pulsars, no pulsations seen in LAT yet  &  \nodata  & \nodata &  msp  & \nmsp \\ 
Pulsar wind nebula  &  PWN  & \nPWN &  pwn  & \npwn \\ 
Supernova remnant  &  SNR  & \nSNR &  snr  & \nsnr \\ 
Supernova remnant / Pulsar wind nebula  &  SPP  & \nSPP &  spp   & \nspp \\ 
Globular cluster  &  GLC  & \nGLC &  glc  & \nglc \\ 
Star-forming region  &  SFR  & \nSFR &  sfr  & \nsfr \\ 
High-mass binary  &  HMB  & \nHMB &  hmb  & \nhmb \\ 
Low-mass binary  &  LMB  & \nLMB &  lmb  & \nlmb \\ 
Binary  &  BIN  & \nBIN &  bin  & \nbin \\ 
Nova  &  NOV  & \nNOV &  nov  & \nnov \\ 
BL Lac type of blazar  &  BLL  & \nBLL &  bll  & \nbll \\ 
FSRQ type of blazar  &    FSRQ  & \nFSRQ &  fsrq  & \nfsrq \\ 
Radio galaxy  &  RDG  & \nRDG &  rdg  & \nrdg \\ 
Non-blazar active galaxy  &  AGN  & \nAGN &  agn  & \nagn \\ 
Steep spectrum radio quasar  &  SSRQ  & \nSSRQ &  ssrq  & \nssrq \\ 
Compact Steep Spectrum radio source  &  CSS  & \nCSS &  css  & \ncss \\ 
Blazar candidate of uncertain type  &  BCU  & \nBCU &  bcu  & \nbcu \\ 
Narrow-line Seyfert 1  &  NLSY1  & \nNLSY &  nlsy1  & \nnlsy \\ 
Seyfert galaxy  &  SEY  & \nSEY &  sey  & \nsey \\ 
Starburst galaxy  &  SBG  & \nSBG &  sbg  & \nsbg \\ 
Normal galaxy (or part)  &  GAL  & \nGAL &  gal  & \ngal \\ 
Unknown  &  UNK  & \nUNK &  unk  & \nunk \\ 
Total  &  \nodata  & \nID &  \nodata  & \nass \\ 
\hline
Unassociated  &  \nodata  & \nIDna &  \nodata  & \nna \\ 
\enddata
\tablecomments{The designation `spp' indicates potential association with SNR or PWN.  Designations shown in capital letters are firm identifications; lower case letters indicate associations. 
}
\end{deluxetable}

A notable addition with respect to 4FGL-DR4 is the improved treatment of associations with star-forming regions (SFRs) and open clusters. Several recent works have convincingly established the case for these objects to be the sites of detectable gamma-ray emission \citep[][and references therein]{Tib21, Per25}. 
We will generically  refer to them as star-forming regions, although some are genuine open clusters. In 4FGL-DR4, three identified (owing to their extension) sources (Westerlund 2, rho Ophiuchi, and Cygnus X)  and three associated SFRs (NGC 346, Sh 2-148, Sh 2-152) were already reported. The classes of W\,3 and Rosette  have been changed to candidate star-forming regions. Except for a few sources described below, in the case of conflicting associations involving SFRs, we have favored the classes other than SFRs to be conservative.  

Known  SFRs are extended (up to several degrees across), numerous (thousands) and closely clustered along the Galactic plane  making the association scheme described above impractical. Applying that scheme to SFRs small enough to be considered as point-like does lead to a small ($\simeq 5$) number of new associations but none of which was missed by the procedure outlined below.   Associations have been performed following different approaches. We used the WISE catalog of H\,{\sc ii} regions \citep{And14}, where only sources of the  ``known'' class and with radii less than $0.3\degr$ were retained. 
This catalog was supplemented  by the list, slightly expanded,  of luminous star-forming complexes of the RMS survey \citep{Urq14}, with the same condition on the radius. In the first approach, some sources classified as ``UNK'', whose associations arose from the radio/X-ray surveys used in the likelihood-ratio method, turn out to be  coincident with SFRs. We have then renamed and reclassified them accordingly. These SFRs are NGC 3603, RCW 38, NGC 6611, Sh 2-87, W 40, and 6 other WISE objects. They all have PROB\_LR$>$0.8 in the FITS file. In the second approach, we have computed the probabilities of association with a WISE/RMS counterpart  for FL16Y sources whose positions fall within the extension of that counterpart. The probability reads:
\begin{equation}  P=(1+\alpha(1-\exp(-\rho S))^{-1}
\end{equation}
where $\rho$ represents the local density of SFRs and $S$ their mean area. $\alpha$ is a constant related to the fraction of false positives, derived from the condition $N_F=\sum (1-P_i)$, where $N_F$ is the number of false coincidences determined by Monte-Carlo simulations. Six sources are associated that way. Finally, we have promoted two potential associations to high-confidence ones, corresponding to RCW 36 and Danks 1, owing to the consideration that they belong to the ``known'' population of gamma-ray star forming regions as defined in \cite{Per25}. In the same vein, we have replaced the  Kleinmann star association of FL16Y J1820.4$-$1609  with NGC 6618 (M 17).

% WISE G006.005-00.368, WISE G035.075-01.519, WISE G043.170-00.004, WISE G048.630+00.230, Sh 2-87, WISE G075.842+00.404, WISE G094.389-05.513,WISE G112.212+00.229.   FL16Y J2256.6+5830 sfr 0.0 0.950803 

%24 MSPs  PSR J0329+5035 PSR J0347-6400 PSR J0447+2447 PSR J0642-6502 PSR J0646-5455 PSR J0739-4527 PSR J0749-4421 PSR J0836-6035  PSR J0838-2827  PSR J0843+6713 PSR J1102+0249 PSR J1207-6900  PSR J1544-2555 PSR J1603-6011  PSR J1624-3951 PSR J1659-1444 PSR J1709-0333 PSR J1752-4450  PSR J1803-4719  PSR J1845+0201 PSR J1909-3744  PSR J1946+3417 PSR J2045-6837  PSR J2051+5050

%5 PSRsPSR J0729-1836 J1410-6132  PSR J1846-0258  PSR J1954+3852 PSR J2016+3711

% Acknowledgments
\begin{acknowledgments}
The \textit{Fermi} LAT Collaboration acknowledges generous ongoing support
from a number of agencies and institutes that have supported both the
development and the operation of the LAT as well as scientific data analysis.
These include the National Aeronautics and Space Administration and the
Department of Energy in the United States, the Commissariat \`a l'Energie Atomique
and the Centre National de la Recherche Scientifique / Institut National de Physique
Nucl\'eaire et de Physique des Particules in France, the Agenzia Spaziale Italiana
and the Istituto Nazionale di Fisica Nucleare in Italy, the Ministry of Education,
Culture, Sports, Science and Technology (MEXT), High Energy Accelerator Research
Organization (KEK) and Japan Aerospace Exploration Agency (JAXA) in Japan, and
the K.~A.~Wallenberg Foundation, the Swedish Research Council and the
Swedish National Space Board in Sweden.
 
Additional support for science analysis during the operations phase is gratefully
acknowledged from the Istituto Nazionale di Astrofisica in Italy and the Centre
National d'\'Etudes Spatiales in France. This work performed in part under DOE
Contract DE-AC02-76SF00515.

This work made extensive use of the ATNF pulsar  catalog\footnote{\url{http://www.atnf.csiro.au/research/pulsar/psrcat}}  \citep{ATNFcatalog}.  This research has made use of the \citet{NED1} which is operated by the Jet Propulsion Laboratory, California Institute of Technology, under contract with the National Aeronautics and Space Administration, of the SIMBAD database \citep{SIMBAD_2000} operated at CDS Strasbourg, France, and of archival data, software, and online services provided by the ASI Science Data Center (ASDC) operated by the Italian Space Agency.
We used the Manitoba SNR catalog \citep{Ferrand2012_SNRCat} to check recently published extended sources. We acknowledge the Einstein@Home project
for providing new pulsar associations through the dedicated efforts of the Einstein@Home volunteers. The Einstein@Home project is supported by the  NSF award 1816904.
\end{acknowledgments}

\software{Gardian \citep{Diffuse2}, GALPROP\footnote{\url{http://galprop.stanford.edu}} \citep{GALPROP17}, HEALPix\footnote{\url{http://healpix.jpl.nasa.gov/}} \citep{Gorski2005}, Aladin\footnote{\url{http://aladin.u-strasbg.fr/}}, TOPCAT\footnote{\url{http://www.star.bristol.ac.uk/\~mbt/topcat/}} \citep{Tay05}}

\facility{\Fermilat}

% Bibliography
\bibliography{Bibtex_5FGL_v1}

\end{document}